\documentclass{elsarticle}

\usepackage{a4wide}
\usepackage{amsfonts}
\usepackage{amsmath}
\usepackage{amssymb}
\usepackage{amsthm,mathtools,url}

\usepackage[numbers]{natbib}
\usepackage[hidelinks]{hyperref}
\usepackage{tikz,pgfplots}
\usepackage{graphicx}
\pgfplotsset{compat=1.3}

\renewcommand{\vec}[1]{\ensuremath \mathbf{\boldsymbol{#1}}}

\newtheorem{theorem}{Theorem}[section]
\newtheorem{lemma}[theorem]{Lemma}
\newtheorem{remark}[theorem]{Remark}
\newtheorem{definition}[theorem]{Definition}
\newtheorem{example}[theorem]{Example}
\newtheorem{corollary}[theorem]{Corollary}

\numberwithin{equation}{section}
\usepackage{subfigure}

\newcommand{\bend}{\hspace*{0ex} \hfill \hbox{\vrule height
    1.5ex\vbox{\hrule width 1.4ex \vskip 1.4ex\hrule  width 1.4ex}\vrule
    height 1.5ex}}

\long\def\symbolfootnote[#1]#2{\begingroup%
  \def\thefootnote{\fnsymbol{footnote}}\footnote[#1]{#2}\endgroup}

\usepackage{ifthen}
\newcounter{todocounter}
\newcommand{\todo}[2][noisnotdefined]{
 \marginpar{\fcolorbox{black}{yellow}{\footnotesize\textbf{todo}}
 \ifthenelse{\equal{#1}{noisnotdefined}}{}{\textcolor{black}{\newline\tiny #1}}}
 \textbf{\ifthenelse{\equal{#2}{.}}
   {\fcolorbox{red}{white}{\textcolor{red}{$\maltese$}}}{{\textcolor{red}{#2}}}}
 \refstepcounter{todocounter}}

\title{The variant graph approach to improved parent grain reconstruction}

\date{\today}

\author[1]{Ralf Hielscher}
\author[2]{Tuomo Nyyssönen\corref{cor1}}
\cortext[cor1]{Corresponding author, contact: tuomo.nyyssonen@swerim.se}
\author[3]{Frank Niessen}
\author[4]{Azdiar A. Gazder}
\address[1]{Chemnitz University of Technology, Department of
  Mathematics, 09107 Chemnitz, Germany}
\address[2]{Swerim AB, 164 40 Kista, Sweden}
\address[3]{Technical University of Denmark, Department of Mechanical Engineering, 2800 Kgs. Lyngby, Denmark}
\address[4]{Electron Microscopy Centre, University of Wollongong, New South Wales 2500, Australia}

\begin{document}

\begin{abstract}
The variant graph is a new, hybrid algorithm that combines the strengths of established global grain graph and local neighbor level voting approaches, while alleviating their shortcomings, to reconstruct parent grains from orientation maps of partially or fully phase-transformed microstructures. The variant graph algorithm is versatile and is capable of reconstructing transformation microstructures from any parent-child combination by clustering together child grains based on a common parent orientation variant. The main advantage of the variant graph over the grain graph is its inherent ability to more accurately detect prior austenite grain boundaries.

A critical examination of Markovian clustering and neighbor level voting as methods to reconstruct prior austenite orientations is first conducted. Following this, the performance of the variant graph algorithm is showcased by reconstructing the prior austenite grains and boundaries from an example low-carbon lath martensite steel microstructure. Programmatic extensions to the variant graph algorithm for specific morphological conditions and the merging of variants with small mutual disorientation angles are also proposed. The accuracy of the reconstruction and the computational performance of the variant graph algorithm is either on-par or outperforms alternate methods for parent grain reconstruction. 

The variant graph algorithm is implemented as a new addition to the functionalities for phase transformation analysis in MTEX 5.8 and is freely available for download by the community.
\end{abstract}


\maketitle

\section{Introduction}
Many alloys undergo partial or complete phase transformation from a metastable parent to a stable child phase when exposed to thermo-mechanical stimuli. Alloy design begins with optimizing the high temperature thermo-mechanical processing regime of the parent microstructure. Here the morphology and crystallographic texture of the parent phase after high temperature processing affects the final microstructure, phase fractions and mechanical properties of the alloy upon transformation during cooling. Alternatively, phase transformation is also an effective means to improve the mechanical properties of alloys by refining the final grain size and crystallographic texture of the final child phase(s) \cite{Niessen2021a,Kundu2007}. In the specific case of partial phase transformation, where some of the parent phase is retained, multi-phase parent-child microstructures with optimal strength-ductility combinations are designed \cite{Edmonds2006}.  Since microstructure analyses are usually undertaken at room temperature, it is generally only possible to characterize the transformed child phase(s), or a combination of remnant parent and child phase(s). Thus, various methods to reconstruct the high temperature parent phase microstructure from experimental phase and orientation maps have been developed \cite{Cayron2006,Miyamoto2010,Germain2012,Niessen2021b}.

A particular difficulty in reconstructing parent grains are alloys in which both parent and child phases exhibit a high degree of crystallographic symmetry. Perhaps the most challenging example in this category is low-carbon lath martensite formed from high-temperature austenite during quenching. Typically, and depending on the operative orientation relationship (OR), up to $24$ individual child orientation variants are generated from a single parent austenite orientation \cite{Nyyssonen2016}. A second difficulty to reconstruction is the high boundary fraction of $\Sigma$3 annealing twins in the parent austenite microstructure. Twinned grains share a mutual (111) plane across the twin boundary, and consequently, the six martensitic variants formed on this particular habit plane (in both twins) are crystallographically close to equivalent \cite{Miyamoto2010}. A single prior austenite orientation may, and often does, have as many as three twins, all sharing variants across twin boundaries. The third major difficulty is that martensite variants with low misorientations form side-by-side as thin laths; bundles of which make up the well-known block structure in martensite \cite{Morito2003}. The low misorientation angle between the laths within a block makes it difficult to obtain crystallographic information from the microstructure at the level of individual martensite variant orientations. Thus, despite a good amount of work towards improving reconstruction algorithms in recent years, the accurate reconstruction of annealing twin boundaries in low-carbon steels at the level of individual variants remains a huge challenge to this day.

Generally speaking, there are two classes of algorithms for parent grain reconstruction that operate on a grain level, namely, grain graph \cite{Brust2021,Nyyssonen2016,GomesdeAraujo2021} and local neighbor voting types, of which the nucleation-growth approach \cite{Cayron2006,Germain2012,Bernier2014} is the most prominent example. Grain graph algorithms consider a graph in which all child grains in a microstructure are represented as nodes. The boundaries between a grain and its neighbors are represented as edges that connect the nodes. By assigning probabilities of belonging to a common parent grain to these edges, clusters of child grains that are likely to belong to the same parent grain are identified and reconstructed. The approach is computationally efficient and considers all grains within the microstructure. A drawback of this method is that it utilizes scalar probabilities to describe the likelihood of having a common parent orientation. These probabilities are extended to higher order neighboring grains without checking whether or not they apply to the same parent orientation variant; thereby leading to the erroneous clustering of child grains in some localized areas \cite{Nyyssonen2016,GomesdeAraujo2021}. As a consequence, varying degrees of additional processing steps are often required following the application of the grain graph approach.

As stated previously, the nucleation-growth approach is the most prominent example of local neighbor level voting algorithms. It identifies nuclei, i.e. local groups of child grains that have a common parent orientation variant, and allows these nuclei to grow into the surrounding parent phase. The determination of the common (and by extension, correct) parent orientation variant is facilitated by a voting mechanism of the participating child grains. The main advantage of this algorithm is that it is relatively robust. A drawback is that some microstructures may not contain enough nuclei to cover all parent orientations. Furthermore, the growth part of the algorithm may lead to the reversion of parent orientation variants that may not be the best fit. In recent work in Ref. \cite{Huang2020}, the growth stage of nuclei is replaced by sectioning the map in square grids and applying a voting algorithm to the grains in each square grid separately to determine parent orientations.

In this study, we introduce a new, hybrid variant graph algorithm to parent grain reconstruction. The algorithm combines the advantages of the established grain graph and nucleation-growth approaches while alleviating their shortcomings. The variant graph enhances and improves on the grain graph algorithm by enabling transitivity conditions to be satisfied, i.e. the product of two edge probabilities agreeing with the probability that the two outer grains belong to the same parent grain. To achieve this, we do not associate each child grain with only one node in the graph but rather, each child grain is associated with one node for each of the potential parent orientation variants. Consequently, the number of nodes associated with a child grain are the number of parent orientation variants allowed by an orientation relationship. 

Iterating via the variant graph approach automatically generates the most likely parent orientation for each grain such that no clustering is required. The use of sparse matrices for the probability graphs and a streamlined implementation of matrix multiplication results in an efficient and robust reconstruction algorithm suitable for large orientation microscopy data sets. The new approach to parent grain reconstruction is implemented in MTEX 5.8 as as a programmatic addition to the previously established phase transformation analysis framework \cite{Niessen2021b} and is freely available to the community.

\section{The example data set: Low-carbon lath martensite steel}
The new, hybrid algorithm is introduced by applying it to a low-carbon lath martensite steel microstructure. The electron backscattering diffraction map consists of 4 million data points at approximately 0.5 $\mu m$ spacing. All processing prior to the application of reconstruction algorithms was undertaken in MTEX 5.8 \cite{BaHiSc10} and entailed: (i) the removal of single-pixel orientation measurements, leaving $83\%$ indexed points, (ii) the defining of a grain map with an angular threshold of $3^{\circ}$, and (iii) the determination of a refined orientation relationship between austenite and martensite using the method described in Ref. \cite{NyIsPeKu16}.

\begin{figure}
    \centering
    \includegraphics[width=0.75\textwidth]{{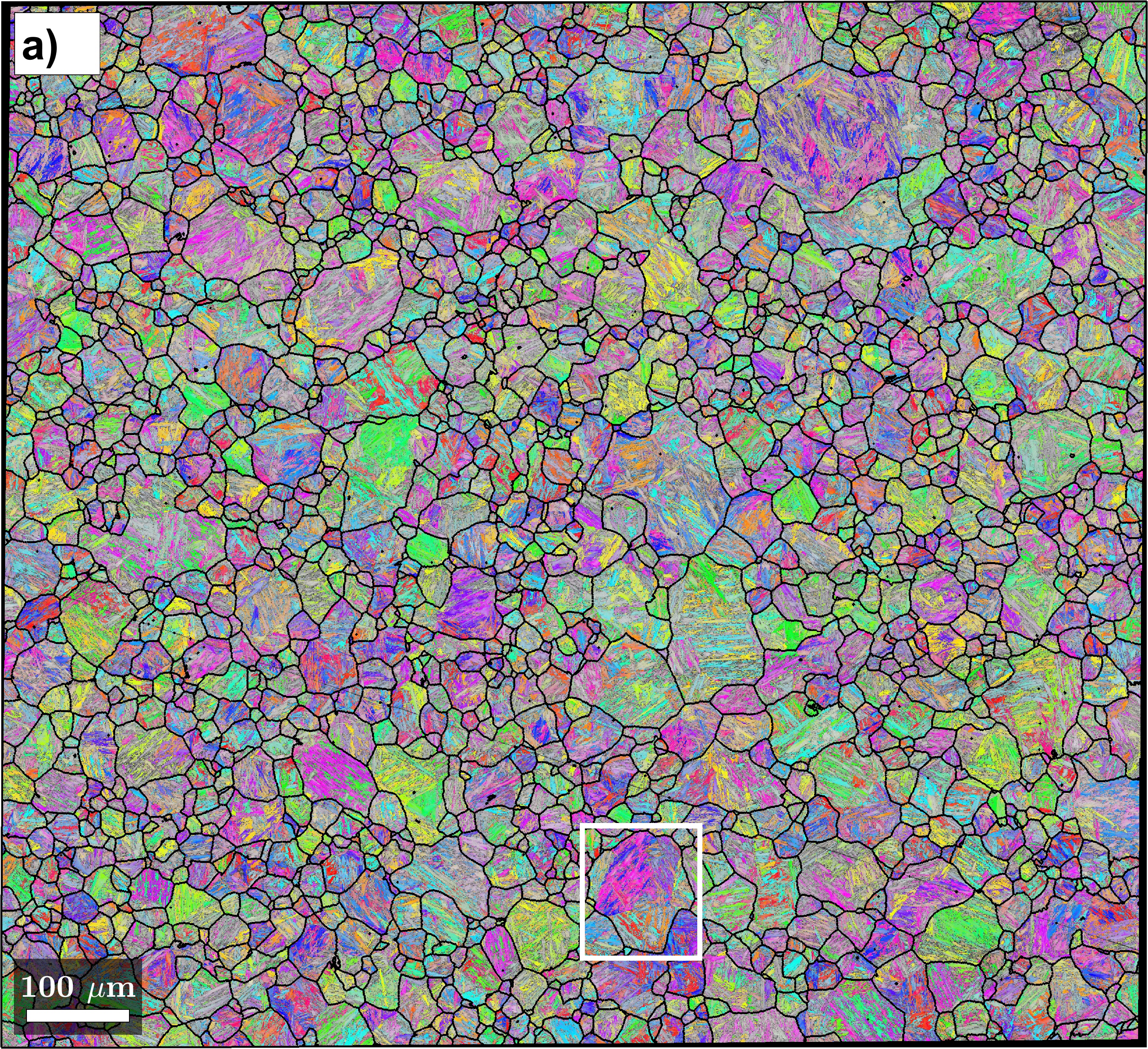}}
    \includegraphics[width=0.66\textwidth]{{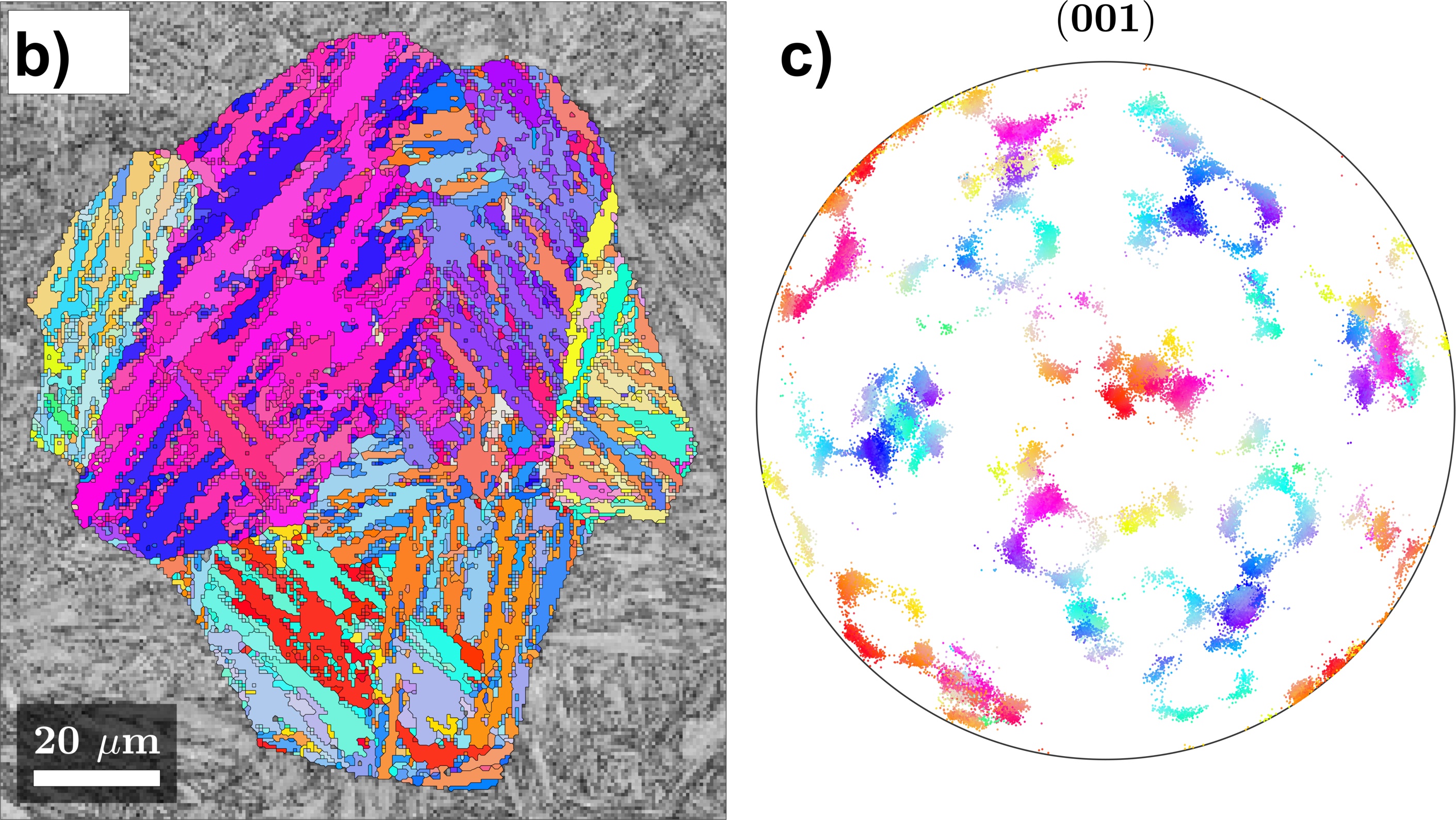}}
    \caption{(a, b) Band contrast maps overlaid with (a) the initial martensite inverse pole figure colors. The reconstructed parent austenite grain boundaries (sans $\Sigma$3 boundaries) are shown in black. (b) A single, heavily twinned parent austenite grain extracted from the map (marked by a white rectangle in (a)). (c) A (001) pole figure showing the martensite orientations in (b).}
    \label{fig:exampleMartensite}
\end{figure}

The map shown in Fig.~\ref{fig:exampleMartensite}(a) comprises band contrast overlaid with individual pixels colored by their inverse pole figure (IPF) orientation. The prior austenite grain boundaries determined by reconstruction in Section \ref{sec:performance} are overlaid on the map as black boundaries to highlight the wide distribution of parent austenite grain sizes. Fig.~\ref{fig:exampleMartensite}(b) highlights a small region of the map containing a single, heavily twinned prior austenite grain. The grain morphology is representative of the hierarchical structure typical of lath martensite \cite{Morito2003}. The (001) pole figure in Fig.~\ref{fig:exampleMartensite}(c) shows the martensite orientations of this grain, illustrated by the same IPF colors.

In subsequent sections, this prior austenite grain is also used as a representative example to review and showcase the performance of the presented reconstruction algorithms. For example, as shown in Section \ref{sec:drawb-reconstr-meth}, there are multiple instances where the martensite variant orientations of twinned austenite grains almost completely overlap. 

\section{Established methods of parent grain reconstruction}
\label{sec:methodology}
In this section we review the two most well established grain level parent grain reconstruction methods. Both methods take advantage of an initial segmentation of the EBSD data into grains $G_{n}$, $n=1,\ldots,N$, cf.~\cite{BaHiSc11}. Each grain is associated with a parent or a child phase and a mean grain orientation $\vec g_{n}$.

Assuming a parent-to-child orientation relationship $\vec g_{\mathrm{p \to c}}$, all potential parent orientation variants of a child orientation $\vec g_{n}$ are given by

\begin{equation}
    \label{eq:parentVariants}
    \vec g_n^j = \vec g_n S_j^c \vec g_{\mathrm{p \to c}},
\end{equation}

where $S_j^c$, $j=1,\ldots,|\mathcal S^c|$ enumerates all child symmetries. In the case of a degenerated orientation relationship, for example, the Nishiyama Wassermann OR \cite{Nishiyama1934}, some of the parent orientation variants may be symmetrically equivalent. To keep the notation simple in such situations, we assume that the index $j$ runs over a maximum set of child symmetries $S_j^c$ such that none of the parent orientation variants $\vec g_n^j$ are symmetrically equivalent.

The goal of parent grain reconstruction is to compute the true parent orientation of each child grain $G_n$ from all possible parent orientations $\vec g_n^j$. The true parent microstructure of a transformed microstructure is revealed by targeted surface etching techniques \cite{Brust2021}. Alternatively, the true parent microstructure of recrystallized austenite reconstructed from individual child grains is generally identified by

\begin{itemize}
    \item Several clusters of neighboring child grains voting for the same parent orientation.
    \item The morphology of a high-temperature parent phase microstructure, typically comprising equiaxed grains.
    \item An annealing twin structure exhibiting a boundary morphology typical of coherent $\Sigma 3$ related twin boundaries  \cite{Mahajan1996}.
\end{itemize}

\subsection{Grain graph based parent grain reconstruction}
\label{sec:graph-based-reconstr}
The method describing the successful identification of clusters of similar parent orientation variants was first introduced by Gomes and Kestens cf.~\cite{GomesdeAraujo2021} and relies on the so-called grain graph. The grain graph is a mathematical description of the adjacency relationships of the grains $G_n$. Each grain $G_n$ corresponds to exactly one node in the graph. Two nodes are connected by an edge if the corresponding grains share a common grain boundary. During the reconstruction algorithm, the edges are labeled with weights $P_{m,n}$ describing the probability that two grains $G_m$ and $G_n$ originate from a common parent grain.  

\begin{figure}
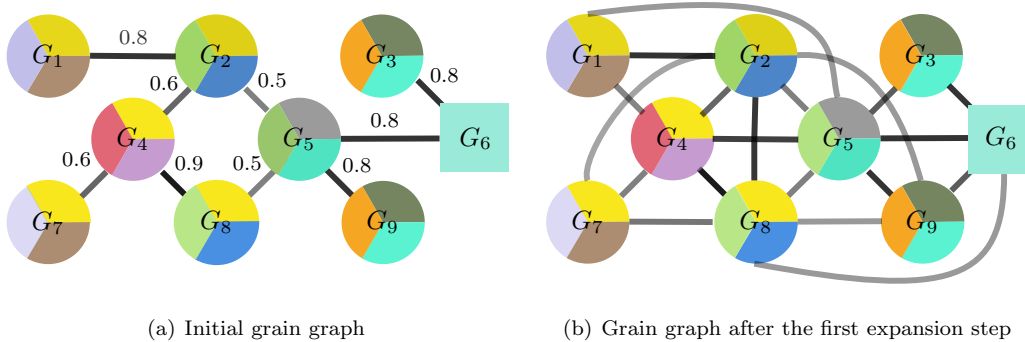

    \centering
    \subfigure[Initial grain graph]{\label{fig:grainGraph}\input{gg1}}
    \subfigure[Grain graph after the first expansion step]{\label{fig:grainGraphSquared}
\input{gg2}}
    \caption{Schematic example of a grain graph. The circular nodes represent child grains $G_1,\ldots,G_9$. Each child orientation is associated with three potential parent variants, each of which are in turn illustrated by differently colored sectors. The square node $G_6$ represents a retained parent grain with a ``cyan'' orientation. 
    In the initial grain graph (a), grains with a common grain boundary are connected by an edge. The darkness or brightness of an edge corresponds to the probability associated with the misorientation angle between best fitting parent variants. For example, grains $G_4$ and $G_8$ are connected by a dark high probability edge since they share a very similar yellow parent orientation. In contrast, grains $G_8$ and $G_5$ are connected by a bright low probability edge since the green parent orientations are slightly different for both child grains.
    Following the first expansion steps of the Markovian clustering algorithm, all second neighbors are then connected by edges. The brightness (and probability) of each of the new edges is computed as the product of the brightness of the individual segments.}
\end{figure}

An example of a grain graph is displayed in Figure \ref{fig:grainGraph}. 
The nodes labeled $G_1,\ldots,G_9$ represent the grains. Adjacent grains are connected by solid lines that are referred to as edges. In order to simplify the example schematic, we assume an OR with only three possible parent variants, i.e., every child grain $G_n$ corresponds to only three potential parent orientations $\vec g_n^j$, $j=1,2,3$. The three possible parent variants of each child grain are visualized by differently colored sectors. 

\tikzset{every picture/.style={line width=0.75pt}} 

Grain $G_6$ is a retained parent grain and, hence, is represented by a square filled with one color. From the colors, the guess is that:

\begin{itemize}
    \item Grains $G_1$, $G_2$, $G_4$, $G_7$ and $G_8$ originate from a common yellow parent grain.
    \item Grains $G_2$, $G_5$ and $G_8$ originate from a common green parent grain.
    \item Grains $G_3, G_5, G_6$ and $G_9$ originate from a common cyan parent grain.
\end{itemize}

Obviously, the three guesses cannot simultaneously be true. The challenging task for any parent grain reconstruction algorithm is to identify the physically most likely solution.

In the case of grain graph based parent grain reconstruction, random walks are simulated through the graph to identify strongly connected clusters. More precisely, the algorithm consists of the following steps:

\begin{enumerate}
\item Constructing the grain graph as described above.
\item Assigning a weight $w_{m,n}$ to each edge which describes the probability that the two adjacent grains $G_n$, $G_m$ originate from a common parent grain.  
\item Clustering the graph into strongly connected components, for example, via the Markovian clustering algorithm \cite{VanDongen2000}.
\item Determining the best fitting parent orientation for each cluster.
\end{enumerate}

The processes of grain graph construction and clustering are detailed further as an understanding of these two items is crucial to the introduction of the new, hybrid variant graph approach in Section \ref{sec:variant-graph-based-reconstr}.

\paragraph{Constructing the grain graph}
The most critical part in constructing the grain graph is the computation of edge weights. Edges are either of \emph{parent - child} or \emph{child - child} types. For a \emph{parent - child} grain pair $G_{m}$ - $G_{n}$, where $\vec g_{m}$ is the parent orientation of $G_m$ and $\vec g_n^j$ are the potential parent orientations of the child grain $G_n$ (see Equ. \eqref{eq:parentVariants}), the misorientation angle $\omega_{m,n}$ between the parent orientation $\vec g_m$ and the best fitting parent variant $\vec g_n^j$ is given by

\begin{equation*}
  \omega_{m,n}
  = \min_{j=1,\ldots,|\mathcal S^c|}
  \omega_p(\vec g_{m},  \vec g_{m,j}).
\end{equation*}

where $\omega_p(\vec g_1,\vec g_2)$ denotes the angular distance in orientation space modulo parent symmetry. 

For two neighboring child grains $G_{m}$, $G_{n}$, the misorientation angle $\omega_{m,n}$ between the best fitting combination of parent variants is

\begin{equation*}
  \omega_{m,n}
  = \min_{i,j = 1,\ldots,|\mathcal S^c|}
  \omega_p(\vec g_m^i,  \vec g_n^j).
\end{equation*}

In order to translate the misfits $\omega_{m,n}$ into probabilities

\begin{equation*}
  P_{m,n} = \Phi(\omega_{m,n})
\end{equation*}

a function $\Phi$ is used that is close to $1$ for very small misfits and decays to zero as the misfit exceeds a certain threshold $\delta$. A common choice for such a function is the Gauss error function $\mathrm{erf}(x)$ scaled by constants $\sigma$ and $\delta$ as

\begin{equation}
  \label{eq:3}
  \Psi(\omega) = 1 - \tfrac{1}{2}\left(1 + \mathrm{erf}(2 \tfrac{\omega - \delta}{\sigma})\right).
\end{equation}

The scaling is chosen such that $\Psi(\delta - \sigma) = 0.75$, $\Psi(\delta) = 0.5$ and $\Psi(\delta + \sigma) = 0.25$. The grain graph is efficiently stored as a sparse $N \times N$ adjacency matrix $P$, where $N$ is the number of grains, containing at row $m$ and column $n$ the entry $P_{m,n}$ if the grains $G_m$ and $G_n$ are connected by an edge with non-zero probability. All remaining matrix entries are zero. 

In Fig.~\ref{fig:grainGraph}, the edge weights are the numbers attached to each of the edges. Furthermore, the darkness of the edges is chosen proportional to the edge weights. In the more complicated follow up graphs, the numbers are omitted.

It should be noted that because of $\omega_{m,n}=\omega_{n,m}$, the matrix $P$ is symmetric and the initial grain graph is not directed. This property is lost during the subsequent clustering step.

\paragraph{Clustering of the grain graph}
While different clustering algorithms may be applied to group grains with a common parent orientation, the following discussion is limited to the commonly used Markovian clustering algorithm \cite{VanDongen2000}. The latter algorithm simulates random walks within a grain graph via the following steps:

\begin{enumerate}
\item Expansion: $P := P \cdot P$\label{item:1}
\item Inflation: $P_{m,n} := P_{m,n}^{\alpha}$\label{item:2}
\item Normalization: $P_{m,n} := P_{m,n} / \sum_{o} P_{m,o}$\label{item:3}
\item Pruning of elements: $P_{m,n} := 0$ if $P_{m,n} < \delta$\label{item:4}
\end{enumerate}

The key assumption of the expansion step \ref{item:1} is as follows: If the probability to walk from node $m$ to node $o$ is $P_{m,o}$ and the probability to walk from node $o$ to node $n$ is $P_{o,n}$, then the probability to walk from $m$ via $o$ to $n$ is $P_{m,o} \cdot P_{o,n} $. Accordingly, summing over all possible intermediate nodes $o$, we obtain the total probability to walk from $m$ to $n$

\begin{equation}
    \label{eq:expGG}
    \tilde P_{m,n} = \sum_{o=1} P_{m,o} P_{o,n}.
\end{equation}

This is exactly the $(m,n)$-th entry of the product matrix $P \cdot P$ in step \ref{item:1}. As the values $\tilde P_{m,n}$ may exceed $1$, they are interpreted as probabilities only after the normalization step \ref{item:3}. 
The purpose of the inflation step \ref{item:2} just before the normalization step is to emphasise higher probability edges over lower probability edges. More precisely, the ratio between two entries in the matrix $P$ becomes more pronounced when higher inflation parameter $\alpha$ values are used. The pruning in the last step \ref{item:4} attempts to keep the matrix $P$ as sparse as possible.

The grain graph Fig.~\ref{fig:grainGraph} after a single run of the steps \ref{item:1}-\ref{item:4} is depicted in Fig.~\ref{fig:grainGraphSquared}. It is observed that by connecting all second order neighbours, the grain graph becomes much more connected. While new edges arise within the yellow, green or cyan clusters, they also unfortunately arise between grains that do not share a common parent orientation. Most notably, the grains $G_4$ and $G_5$ get connected with a high probability edge even though they cannot possibly originate from a common parent grain. This occurs because $G_4$ is connected to $G_5$ by two routes through $G_2$ and $G_8$, respectively. Although both routes contain low probability edges, they sum up to a high probability edge between $G_4$ and $G_5$. As a consequence, all nodes in our example appear as a single cluster. A complex situation as described above is one of the core reasons behind why the Markovian clustering algorithm sometimes generates very large clusters of child grains that do not necessarily agree on a common parent orientation. 

By iterating this process, i.e., by considering $P^{2}$, $P^{4}$, $P^{8}$, up to second, fourth, eight order neighbors are included into the probability matrix. As a consequence, the matrix $P$ includes more and more non-zero elements and, hence, becomes less sparse. The combination of the inflation and normalization steps ensures that during the iteration process for each node, higher probability edges become stronger while lower probability edges decay to zero. This keeps the matrix sparse and eventually ensures that the matrix separates into disconnected components by converging to an idempotent matrix with all entries being either $0$ or $1$.

\paragraph{Computing parent orientations}
After the grain graph converges to an idempotent binary matrix, parent grains are constructed by merging all pairs of child grains $G_m$, $G_n$ with $P_{m,n}=1$. The parent orientation $\vec g^p$ of a grain merged from child grains $G_{m_1},G_{m_2},\ldots,G_{m_X}$
is usually computed by minimizing the mean misorientation angle to the best fitting potential parent orientations $\vec g_{m_k}^j$ of the child grains. This is written as the optimization problem

\begin{equation*}
    \min_{\vec g^p} \sum_{k=1}^X 
    \min_j \omega_p(\vec g^p, \vec g^j_{m_k}).
\end{equation*}

Unfortunately, the grain graph does not contain the information about the best fitting parent orientations $\vec g_{m_k}^j$. This makes the optimization problem computationally expensive to solve. Usually, this step is the most time consuming in the grain graph approach to parent grain reconstruction.

\subsection{Local neighbor level voting based parent grain reconstruction}
\label{sec:vote-based-reconstr}
In local neighbor level voting based reconstruction algorithms, votes for each potential parent orientation are collected from the neighboring grains of each child grain. In the schematic example in Fig.~\ref{fig:grainGraph}, grain $G_2$ collects a yellow vote from $G_1$ and a green vote from $G_3$ while $G_3$ collects a yellow vote from $G_2$ and two cyan votes from $G_4$ and $G_5$. 

In order to formalize the voting, the misorientation angles between potential parent orientation variants of the child grains are considered. If a grain pair comprising a neighboring child grain $G_m$ and a parent grain $G_n$ is considered, the  misorientation angles $\omega_{m,n}^{i}$ between the potential parent orientations $\vec g_m^i$ of $G_m$ and the parent orientation $\vec g_n$ of $G_n$ is denoted by

\begin{equation*}
    \omega_{m,n}^{i} 
    = \omega_p(\vec g_m^i,\vec g_{n})
\end{equation*}

Similar to the grain graph approach, we use a function $\Psi$ to transform the misorientation angles $\omega_{m,n}^{i} $ into values

\begin{equation}
    \label{eq:P2}
    P_{m,n}^{i} = \Psi(\omega_{m,n}^{i})
\end{equation}

that approach zero if the misorientation angle exceeds a certain threshold or are close to one for small misorientation angles. The value $P_{m,n}^{i}$ is interpreted as the voting weight for the parent orientation $\vec g_m^i$ of the child grain $G_m$. In this sense, the neighboring parent grain $G_n$ generates a voting weight $P_{m,n}^{i}$ for each potential parent orientation $\vec g_m^i$ of the child grain $G_m$. Depending on the threshold of the function $\Psi$, most of these weights will be zero.

In the case of a pair $G_m$, $G_n$ of neighboring child grains, we have $|\mathcal S^c| \times |\mathcal S^c|$ misorientation angles

\begin{equation*}
    \omega_{m,n}^{i,j} 
    = \omega_p(\vec g_m^i,\vec g_n^j)
\end{equation*}

between all possible combinations of potential parent orientations $\vec g_m^i$ of $G_m$ and potential parent orientations $\vec g_n^j$ of $G_n$. Again, the function $\Psi$ is used to transform the misorientation angles into voting weights. More specifically, for a potential parent orientation $\vec g_m^i$ of grain $G_m$, the grain $G_n$ generates a vote with the weight 

\begin{equation*}
    P_{m,n}^{i} = \max_j \Psi(\omega_{m,n}^{i,j}).
\end{equation*}

Depending on the threshold angle of the function $\Psi$ and the orientation relationship between the two child orientations $\vec g_m$ and $\vec g_n$, most of the voting weights will be zero.

By summing the voting weights $P_{m,n}^i$ separately for each potential parent orientation $\vec g_m^i$ of $G_m$ over all neighboring grains $G_n$, we obtain the final voting weights

\begin{equation}
\label{eq:PmiVote}
  P_{m}^{i}
  = \sum\limits_{n} P_{m,n}^{i}.
\end{equation}

If it is momentarily assumed that the function $\Psi$ is simply a cutoff function, i.e., $\Psi(\omega)=0$ if $\omega$ is above a certain threshold and $\Psi(\omega)=1$ otherwise,  then $P_{m}^{i}$ simply counts the number of neighboring grains that have a potential parent orientation close to $\vec g_m^i$. In order to make the voting weights comparable across different child grains, the voting weights are often normalized by

\begin{equation*}
  \bar P_{m}^{i}
  = \frac{P_{m}^{i}}{\sum\limits_{i} P_{m}^{i}}
\end{equation*}

such that the sum of all votes for any child grain $G_m$ satisfies $\sum\limits_{i} \bar P_{m}^{i}=1$.

In its simplest form, a voting algorithm requires the selection of the parent orientation $\vec g_m^i$ with the highest voting weight $\bar P_{m}^{i}$ for each child grain $G_m$. More involved schemes may assign parent orientations to only those child grains when the highest voting weight exceeds a certain threshold, or when the difference between the highest and second highest voting weight is sufficiently large. After parent orientation assignment, the process is iterated with equal or relaxed criteria. In fully transformed microstructures, the iteration process is often started by considering only child - child grain pairs, which is termed as the nucleation step, followed by several growth steps where only parent - child grain pairs are considered.

\subsection{Advantages and drawbacks of established parent grain reconstruction methods}
\label{sec:drawb-reconstr-meth}
The global grain graph and the local neighbor level voting approaches to parent grain reconstruction described in the above two sub-sections have complementary advantages and limitations. While the grain graph algorithm considers only the best fitting combination of parent orientations for each pair of neighboring grains, the voting based algorithm considers all possible parent orientations for a child grain and accordingly, all possible fits to the neighboring grains. Furthermore, while the grain graph approach also considers higher-order neighbors, i.e. grains that are further away, the voting based approach only considers first-order neighbors and hence makes its choice based on local and immediate neighbor information. 

The aim of the new and hybrid variant graph approach is to combine the advantages of the two approaches while overcoming their shortcomings. In this regard, Section \ref{sec:graph-based-reconstr} refers to one of the central assumptions of the conventional Markovian clustering algorithm which is as follows.

Consider that $P_{m,o}$ is the probability that the two child grains $G_m$ and $G_o$ belong to the same parent grain and $P_{n,o}$ is the probability that the child grains $G_n$ and $G_o$ belong to the same parent grain. In this case, the product $P_{m,o} \cdot P_{o,n}$ is the probability that the grains $G_m$ and $G_n$ belong to the same parent grain. Generally speaking, this assumption is not true. It may happen that $G_m$ and $G_o$ agree on a common parent orientation which may be completely different to the parent orientation that $G_n$ and $G_o$ agree on. In such a situation, the true probability $P_{m,n}$ is much lower than $P_{m,o} \cdot P_{o,n}$. The consequence of this condition not being satisfied is frequently observed when the conventional Markovian clustering algorithm detects potential parent grains where no reasonable grain orientation is assigned \cite{Niessen2021b}.

\begin{figure}
    \centering
    \includegraphics[width=1\textwidth]{{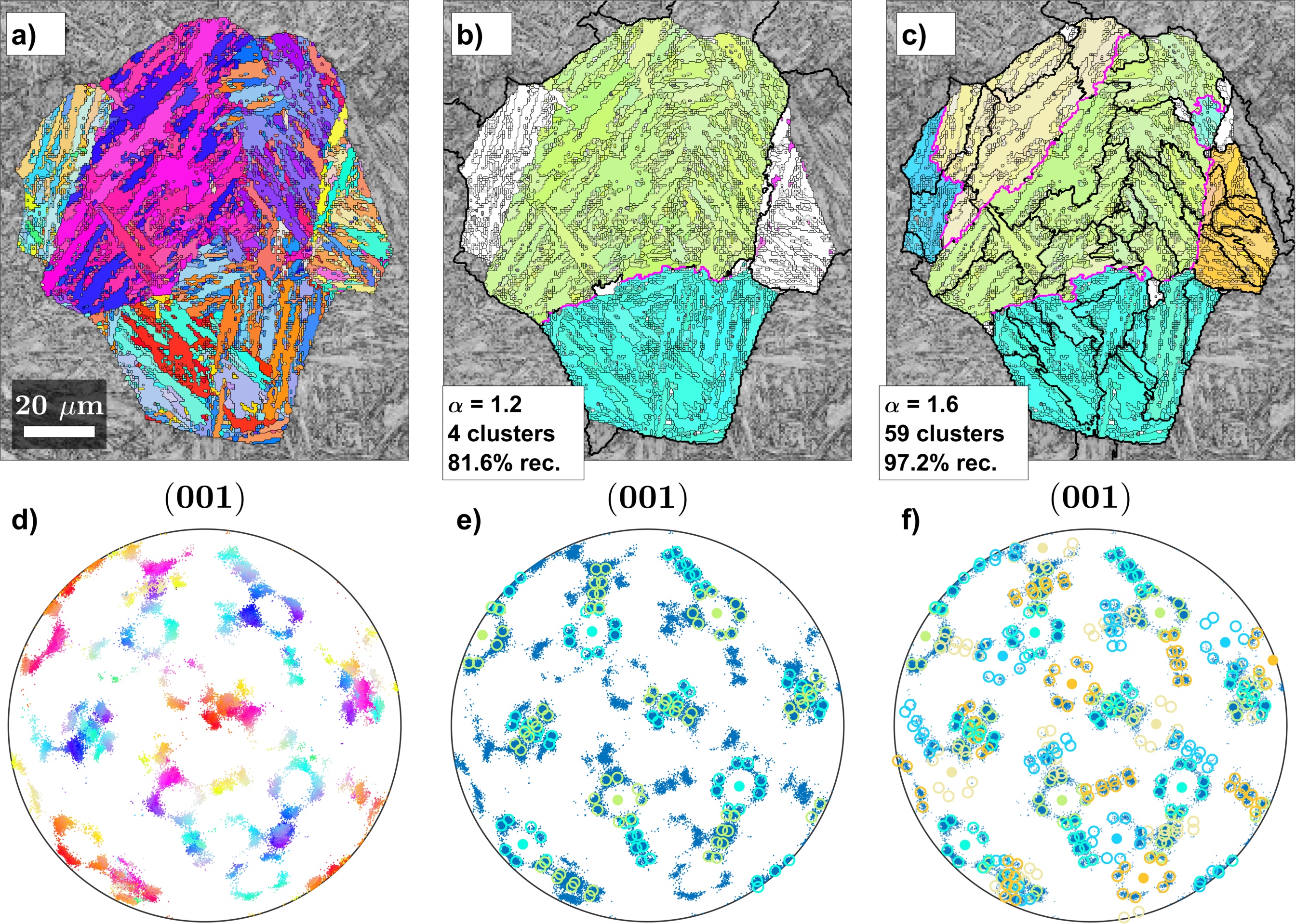}}
    \caption{(a, b, c) Band contrast maps overlaid with (a) the initial martensite inverse pole figure colors for a single, heavily twinned parent austenite grain. The parent austenite grain map reconstructed using the grain graph algorithm after clustering with (b) $\alpha$ = 1.2 and (c) $\alpha$ = 1.6. (d, e, f) (001) pole figures showing (d) martensite orientations and (e, f) grain graph based parent austenite orientations with calculated martensite orientations overlaid on to measured martensite orientations for (e) $\alpha$ = 1.2 and (f) $\alpha$ = 1.6.}
    \label{fig:graingraphexample}
\end{figure}

This issue is highlighted by showing the result of applying Markovian clustering to segment the grain graph generated from the example data set in Fig.~\ref{fig:graingraphexample}. The edge weights for the grain graph were determined by Equ. \eqref{eq:3} using the parameters $\delta$ = 5\textdegree and  $\sigma$ = 1.5\textdegree. The graph was then clustered in two separate calculations, first with the inflation parameter $\alpha$ = 1.2 and then with $\alpha$ = 1.6. After convergence, the parent orientations were computed in a separate step as outlined in Section \ref{sec:graph-based-reconstr}. Grains that were successfully reconstructed to a parent orientation are shown using inverse pole figure colors while the grains for which no parent orientation was found are shown in white. The clusters formed by the algorithm are marked by thick black boundaries.

Fig.~\ref{fig:graingraphexample}(b) clearly shows the effect of using a very low value for the inflation parameter. Under-clustering results in several clusters that extend beyond the boundaries of the parent austenite grain and the determined orientations are insufficient to account for the observed martensite orientations. The latter is also as shown by the pole figure below in (Fig.~\ref{fig:graingraphexample}(e)). Conversely, using a larger inflation parameter results in over-clustering of the parent austenite structure (Fig.~\ref{fig:graingraphexample}(c)). As opposed to Fig.~\ref{fig:graingraphexample}(b), over-clustering enables the calculation of a larger number of parent orientations with more than 97\% of child grains undergoing reconstruction. The latter also account for the observed martensite orientations in Fig.~\ref{fig:graingraphexample}(e). However, the biggest disadvantage to over-clustering is the increase in the number of optimization problems that need to be resolved in the second reconstruction step. In turn, this issue increases the amount of computational resources needed for reconstruction. A second consideration is that extreme over-clustering results in clusters comprising one or two grains; resulting in poorly defined parent orientations.

\begin{figure}
    \centering
    \includegraphics[width=0.66\textwidth]{{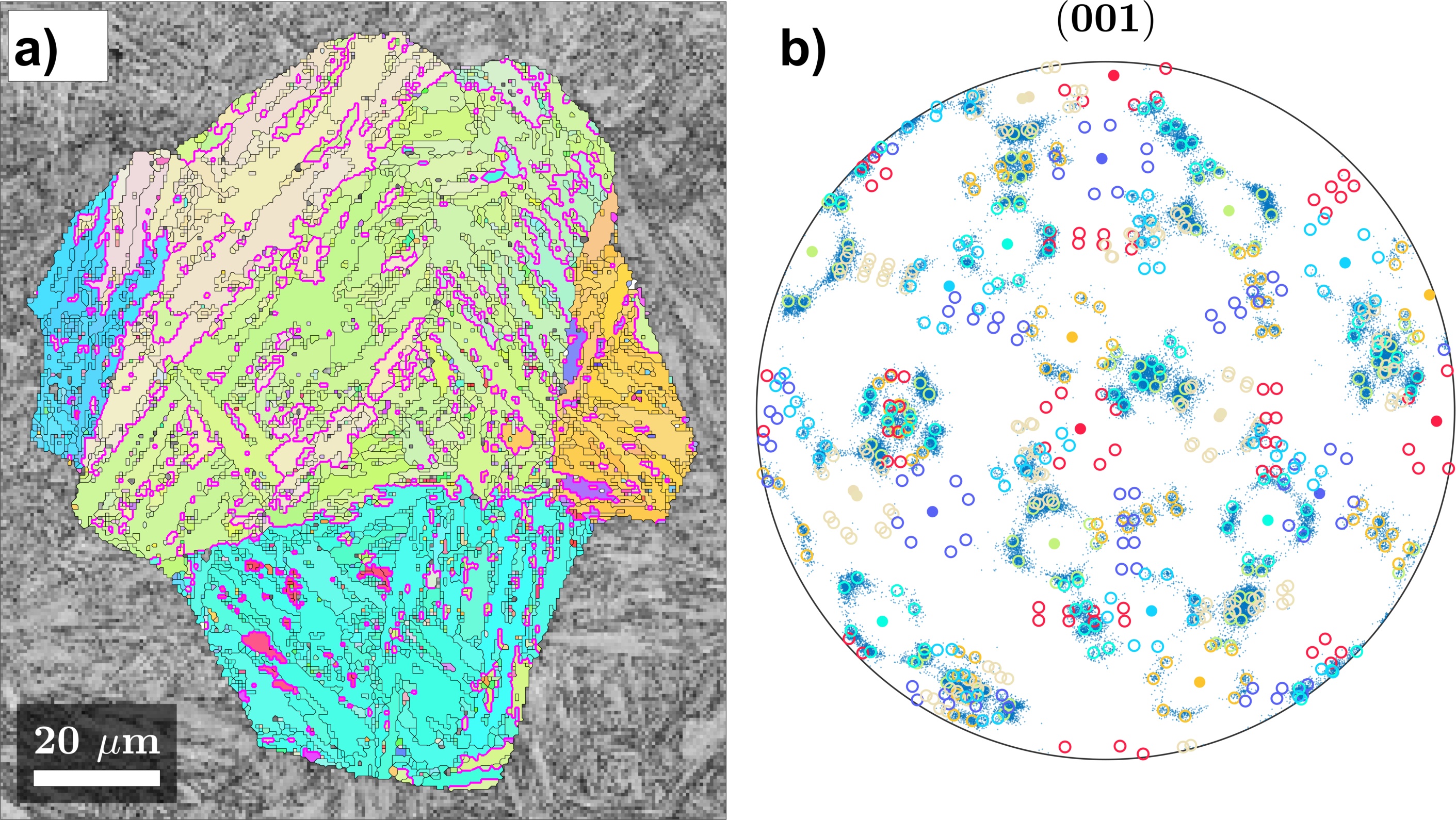}}
    \caption{(a) Band contrast map overlaid with the parent austenite grain map reconstructed using the voting algorithm. (b) (001) pole figure showing martensite orientations, the reconstructed austenite orientations and the theoretical martensite orientations calculated from these.}
    \label{fig:votingexample}
\end{figure}

On the other hand, local neighbor level voting based reconstruction applies variant level information between neighboring grain pairs; thus ensuring a locally viable solution for each reconstructed parent austenite grain. However, this locally viable solution is reached after applying either: (i) a nominal orientation relationship referenced from the literature, or (ii) an average orientation relationship determined for the entire data set. Previous work has shown that there is considerable local variation in the orientation relationship between austenite and martensite \cite{Cayron2014,Sandvik1983} such that applying a representative average orientation relationship could result in the local misindexing of parent austenite variants.

Fig.~\ref{fig:votingexample} shows the result of reconstruction using the voting algorithm. For each child grain a parent orientation was determined based on the votes of neighboring child grains, as outlined in Section \ref{sec:vote-based-reconstr}. The voting based reconstruction returned a high fraction of local twinned orientations (pink boundaries), seen as inclusions within larger grains. In addition, although several parent austenite orientations are indexed as they produce locally viable solutions, they do not correspond well with the observed martensite orientations. This aspect is clearly seen in the pole figure in Fig.~\ref{fig:votingexample}(b). 

In summary, while the grain graph algorithm is able to efficiently cluster the data into segments, it is prone to under and over -clustering as it lacks information at the individual parent variants level. Meanwhile, applying variant level information locally in voting based algorithms results in frequent misindexing; especially when nominal or averaged orientation relationship are used. As detailed next in Section \ref{sec:variant-graph-based-reconstr}, the new, hybrid variant graph approach eliminated the shortcomings identified in these two algorithms by combining variant level information with larger-scale clustering.

\section{Variant graph based parent grain reconstruction}
\label{sec:variant-graph-based-reconstr}

\subsection{The variant graph}
The variant graph is a generalization of the grain graph. In the grain graph each child grain is represented by a single node $G_m$. Alternatively, the variant graph contains as many nodes $\vec g_m^i$ as the number of potential parent orientation variants allowed by an orientation relationship for each child grain $G_m$. In both methods, the parent grains appear as single nodes $\vec g_m$. 

Two nodes in the variant graph are connected by an edge if two conditions are met: (i) they correspond to adjacent grains, and (ii) their misorientation angle is below a certain threshold. The weighting of these edges is computed from the misorientation angle between the node orientations using the function $\Psi$, c.f. Equ. \eqref{eq:3}. More precisely, for two adjacent child nodes $\vec g_m^i$, $\vec g_n^j$, the edge weight is 

\begin{equation*}
  P_{m,n}^{i,j} = \Psi(\omega_p(\vec g_m^i, \vec g_n^j))   
\end{equation*}

and for a pair $\vec g_m^i$, $\vec g_{n}$ of adjacent child - parent nodes, the edge weight is

\begin{equation*}
  P_{m,n}^{i} = \Psi(\omega_p(\vec g_m^i, \vec g_n)).   
\end{equation*}

\begin{figure}
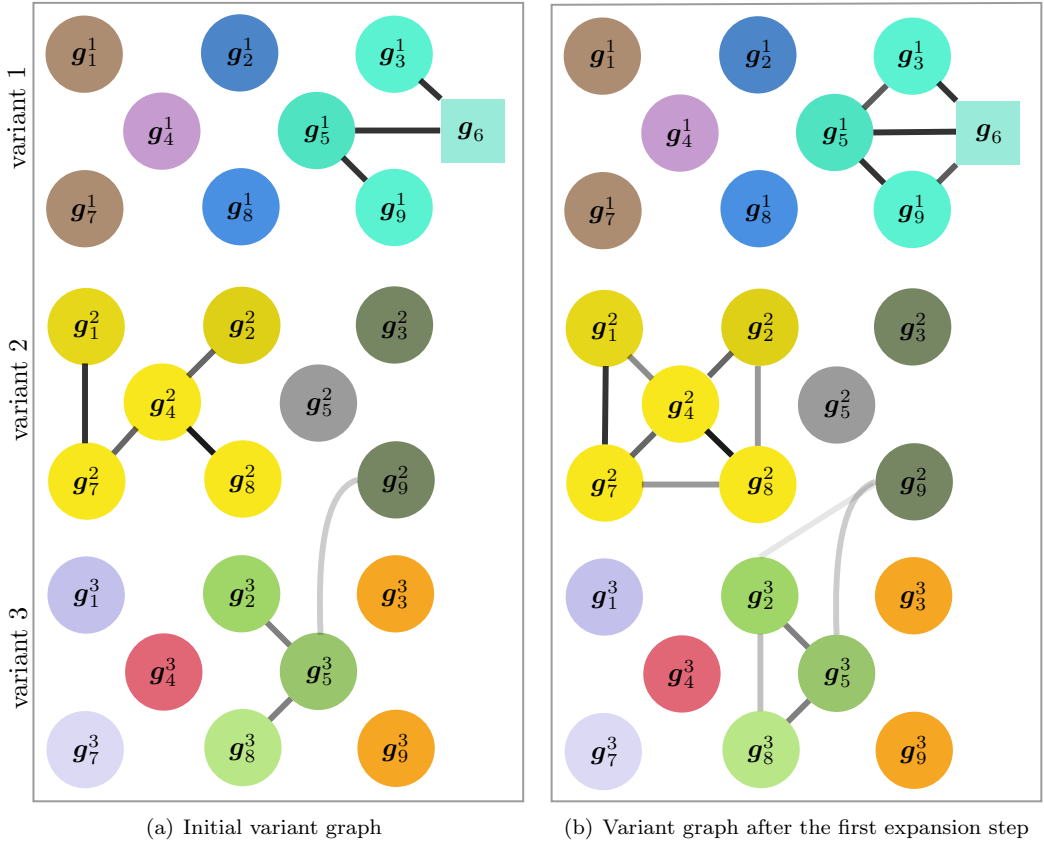

    \centering
    \subfigure[Initial variant graph]{\label{fig:variantGraph}\input{vg1}}
    \subfigure[Variant graph after the first expansion step]{\label{fig:variantGraph2}\input{vg2}}
    \caption{The variant graph corresponding to the grain graph in Fig.~\ref{fig:grainGraph}. Each child grain $G_n$ is represented by three nodes $\vec g_n^1$, $\vec g_n^2$ and $\vec g_n^3$ which refer to the three potential parent orientations. An exception is the square parent grain $G_6$ which appears as a single parent orientation $\vec g_6$. In Fig.~\ref{fig:variantGraph} the nodes are connected by an edge if the child grains are adjacent and the misorientation angle between the potential parent orientations is below a certain threshold. Edges are not restricted within the same variant number as the edge between the orientations $\vec g_5^3$ and $\vec g_9^2$ illustrates. Fig.~\ref{fig:variantGraph2} displays the variant graph after the first expansion step.}
\end{figure}

The variant graph for the example in Fig.~\ref{fig:grainGraph} is shown in Fig.~\ref{fig:variantGraph}. Since this simplified example only considers three parent variants, every child grain appears as three nodes $\vec g_n^1$, $\vec g_n^2$ and $\vec g_n^3$. Meanwhile, the number of edges has only increased by one. The additional edge stems from the two edges $\vec g_5^1 - \vec g_9^2$ and $\vec g_9^3 - \vec g_5^3$ connecting the grains $G_5$ and $G_9$. Using the variant graph approach, the clusters of green, yellow and cyan parent orientations are far more separated compared to the grain graph.

Here the crucial distinction in the theories behind the grain and variant graph approaches is emphasized:

\begin{description}
\item[Grain graph:] In the grain graph approach, two adjacent grains $G_{m}$ and $G_{n}$ are connected by an edge with a weight $P_{m,n}$ if there is a pair of parent orientations $\vec g_{m}^{i}$, $\vec g_{n}^{j}$ among all possible parent orientations of grain $G_{m}$ and all possible parent orientations of grain $G_{n}$ such that the misorientation angle $\omega_{m,n}^{i,j}=\omega_p(\vec g_m^i, \vec g_n^j)$ is below a certain threshold. However, the critical information on the specific best fitting pair $i, j$ is not stored.

\item[Variant graph:] In the variant graph approach, all possible parent orientations $\vec g_m^i$ for each child grain $G_m$ are stored. Two possible parent orientations $\vec g_m^i$ and $\vec g_n^j$ are connected by an edge with a weight $P_{m,n}^{i,j}$ if the corresponding grains $G_{m}$ and $G_{n}$ are adjacent and the misorientation angle $\omega_p(\vec g_m^i, \vec g_n^j)$ is below a certain threshold.
\end{description}

In a sense, the variant graph is a generalization of the grain graph such that the latter may be directly derived from the former by collapsing the many variant nodes $\vec g_m^i$ into single nodes $G_m$ and by replacing the edge probabilities $P_{m,n}^{i,j}$ between the variant nodes with their maximum 

\begin{equation*}
  P_{m,n} = \max_{i,j} P_{m,n}^{i,j} 
\end{equation*}

over all possible variant pairs which in turn, gives the edge probabilities of the grain graph.

At first glance, the variant graph may appear to be much larger than the grain graph, and hence, may be construed as numerically unwieldy to work with. However, the number of stored edges in the variant graph are of similar magnitude as that for the grain graph approach. More specifically, if the misorientation angle threshold is chosen such that for any pair of neighboring grains $G_m$, $G_n$ only the best fitting pair $\vec g_m^i$, $\vec g_n^j$ of parent orientations is connected by an edge in the variant graph, then the number of edges in the variant graph is equal to the number of edges in the grain graph. Since the amount of memory required to store a graph by a sparse adjacency matrix only depends on the number of edges, it is safely concluded that the variant graph is computationally manageable. 

One of the main advantages of the variant graph is that information about second and third best fits is utilized as well. The inclusion of higher order fits significantly improves the quality of the reconstructed parent grains. However, this comes at a price, with higher memory usage requirements that scale linearly with the number of considered fits. 

Similar to the grain graph, the variant graph is symmetric at the beginning as $P_{m,n}^{i,j} = P_{n,m}^{j,i}$ but this property is lost during the clustering process.

\subsection{Generalization of the Markovian clustering algorithm}
\label{sec:general-markovian}
The idea behind the Markovian clustering algorithm is to compute the probabilities of random walks throughout the grain graph. Therefore, two steps are crucial: (i) the expansion step \ref{item:1}, and (ii) the normalization step \ref{item:3}. These two steps are applied to the variant graph in a straightforward manner.  

In the expansion step \ref{item:1} of the grain graph, a probability $\tilde P_{m,n}$ that two grains $G_m$, $G_n$ belong to a common parent grain is computed by summing the product of such probabilities $P_{m,o} P_{o,n}$ with respect to all middle grains $G_o$. Alternatively, in the variant graph, only those products of probabilities $P_{m,o}^{i,k} P_{o,n}^{k,i}$ that agree with the parent orientation $\vec g_o^k$ of the middle grain are summed up. Since all middle grains and all parent variants need consideration, the sum enlarges to 

\begin{equation}
  \label{eq:expVG}
  \tilde P_{m,n}^{i,j}
  = \sum_{o} \sum_{k}
   P_{m,o}^{i,k} P_{o,n}^{k,j}.
\end{equation}

It should be noted that although two summations indices in Equ. \eqref{eq:expVG} are included, the expansion step is simply the matrix product $\tilde P = P \cdot P$.

In order to turn the expanded matrix $P^2$ again into a probability matrix, the normalization step \ref{item:3} of the conventional Markovian clustering algorithm is required. For the variant graph this is

\begin{equation*}
  \tilde P_{m,n}^{i,j}
  = P_{m,n}^{i,j} / \sum_{i} \bar P_m^i
  \text{ with }
  \bar P_{m}^{i}
  = \sum_n \sum_j P_{m,n}^{i,j}.
\end{equation*}

The normalization ensures that for any grain $G_m$, the total sum of probabilities for its potential parent orientations and all neighbouring potential parent orientations is one, i.e.,

\begin{equation*}
  \sum_n \sum_{i,j} \tilde P_{m,n}^{i,j} = 1.
\end{equation*}

The purpose of the remaining two steps of the Markovian clustering algorithm, (iii) the inflation step \ref{item:2}, and (iv) the pruning step \ref{item:4}, is the keep the grain graph sparse and enforce its convergence to an idempotent matrix. Both steps are generalized to the variant graph as 

\begin{equation*}
  P_{m,n}^{i,j} \leftarrow \left(P_{m,n}^{i,j}\right)^{\alpha}
\end{equation*}

and

\begin{equation*}
  P_{m,n}^{i,j} \leftarrow 0 \text{ if } P_{m,n}^{i,j}<\delta.
\end{equation*}

\subsection{Computing parent orientations}
\label{sec:compute-parent-or}
Unlike the grain graph approach, no optimization problems need to be resolved in order to determine parent orientations. In fact, once the generalized Markovian clustering algorithm has converged, the edge probabilities $P_{m_0,n}^{i,j}$ for a certain child grain $G_{m_0}$ are all zero except for a single entry $P_{m_0,n^*}^{i^*,j^*} = 1$. This single edge indicates that the grains $G_{m_0}$ and $G_{n^*}$ belong to a common parent grain with the parent orientation of $G_{m_0}$ being $\vec g_{m_0}^{i^*}$ and the parent orientation of $G_{n^*}$ being $\vec g_{n^*}^{j^*}$.

In fact, even the requirement of full convergence for the generalized Markovian clustering may be relaxed. In this case, for all possible parent orientations $\vec g_m^i$ of a fixed child grain $G_m$, the most robust solution is to compute the sum of the edge probabilities to all other grains, i.e.,

\begin{equation*}
  \bar P_{m}^{i}
  = \sum_n \sum _j P_{m,n}^{i,j}
\end{equation*}

and select the parent orientation $\vec g_m^i$ corresponding to the largest value of $\bar P_{m}^{i}$. When normalized to one by setting

\begin{equation}
  \label{eq:Pmi}
  P_{m}^{i}
  = \bar P_{m}^i / \sum_{i} \bar P_m^i
\end{equation}

the values are interpreted as probabilities voting for the parent orientation $\vec g_m^i$ of the child grain $G_m$. In this respect, the variant graph directly resembles the voting based parent grain reconstruction approach from Section \ref{sec:vote-based-reconstr}.

This final step of parent orientation determination is much faster than the corresponding step in the grain graph approach.

\subsection{Variant graph clustering of the example data set}
\label{sec:example-applications}
\begin{figure}
    \centering
    \includegraphics[width=1\textwidth]{{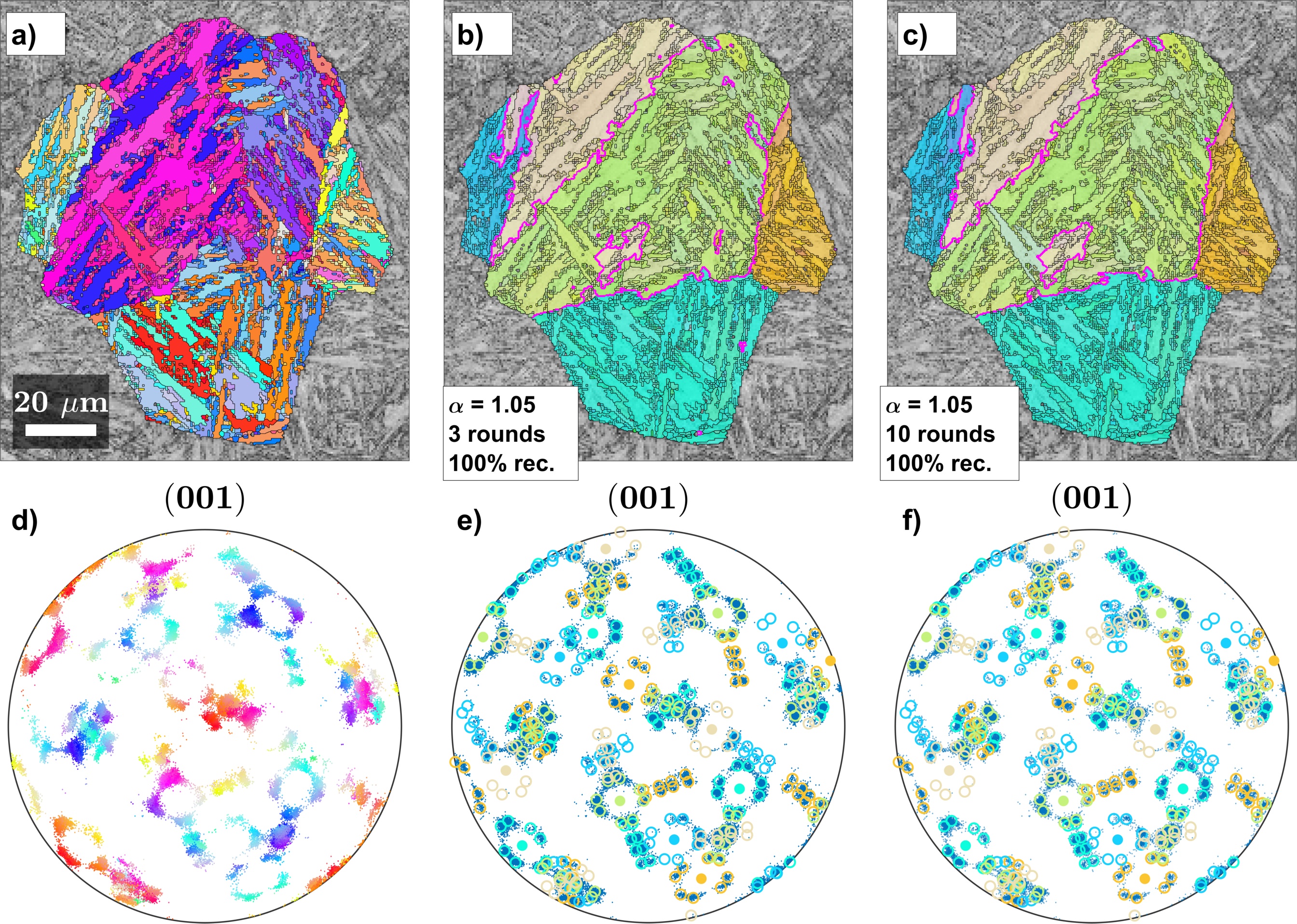}}
    \caption{(a, b, c) Band contrast maps overlaid with (a) the initial martensite grain map for a single, heavily twinned parent austenite grain and parent austenite grain map reconstructed with the variant graph algorithm after (b) 3 and (c) 10 iterations, along with (d, e, f) the (001) pole figures showing (d) martensite orientations and parent austenite orientations after (e) 3 and (f) 10 iterations with calculated martensite orientations overlaid on measured martensite orientations.}
    \label{fig:martensiteOrientations}
\end{figure}

 Fig.~\ref{fig:martensiteOrientations} shows the results of parent grain reconstruction using the variant graph algorithm, along with the reconstructed orientations and calculated martensite variants on (001) pole figures. The edge weights for the variant graph were determined in a similar manner as the grain graph using Equ. \eqref{eq:3} and the parameters $\delta$ = 5\textdegree and  $\sigma$ = 1.5\textdegree. The variant graph was then run through the Markovian clustering algorithm using an inflation parameter $\alpha$ = 1.05. The graph was used to reconstruct parent austenite orientations after 3 and 10 iterations.
 
 Using the variant graph algorithm, all the grains in the original map have been successfully assigned a parent austenite orientation. In addition, there is no spilling over of clusters to neighboring parent grains as seen when using the grain graph in Fig.~\ref{fig:graingraphexample}(b). Since the variant graph stores the orientations as well as the probability value for each edge, there is no chance of such spillover. Inspection of the pole figure indicates that the variants calculated from the reconstructed parent orientations account for each martensite orientation in the original grain map after only 3 iterations. Fig.~\ref{fig:martensiteOrientations}(c) shows the result of parent grain reconstruction after 7 more iterations, i.e., for a total of 10 iterations. At this point, all child grains are assigned parent austenite orientations and all martensite orientations are accounted for. There is, however, a slight difference in the regions assigned to the various parent orientations.

\begin{figure}
    \centering
    \includegraphics[width=1\textwidth]{{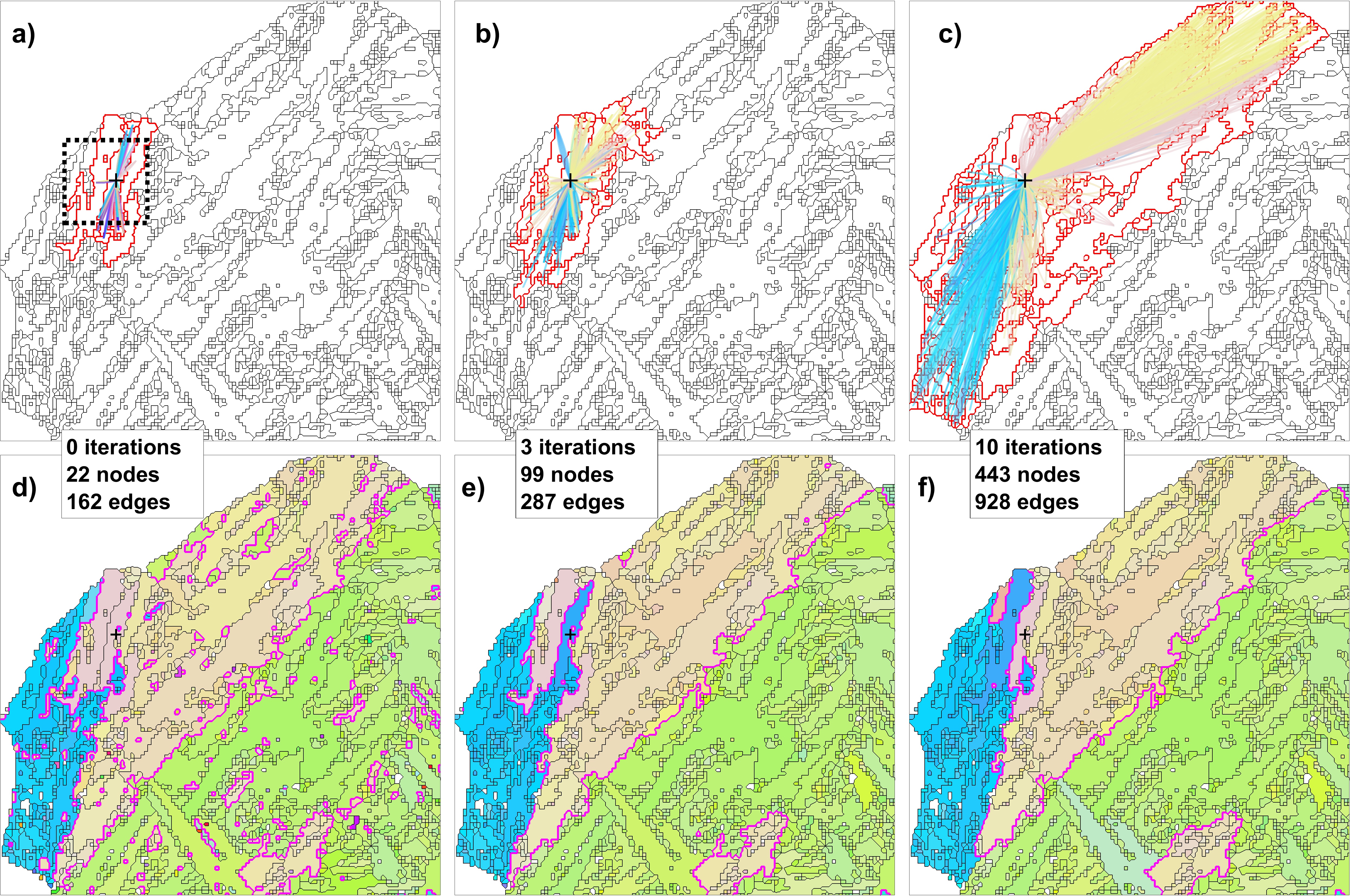}}
    \caption{(a, b, c) Edges and nodes in a variant graph for a selected martensite grain and (d, e, f) the parent orientations from the variant graph algorithm after (a, d) 0, (b, e) 3 and (c, f) 10 iterations. The edges in the top image comprise the IPF colors after the parent orientation of each edge. The area marked with a dashed line in (a) is looked at more closely in Fig. \ref{fig:merge_variants}.}
    \label{fig:variantGraphIterations}
\end{figure}

The progress of the variant graph algorithm for a selected grain in the martensite grain map is shown in Fig.~\ref{fig:variantGraphIterations}. The upper figure shows a selected grain in the upper left-hand region of the prior austenite grain shown in ~\ref{fig:martensiteOrientations}, as well as the neighboring grains for which variant pairs are found that meet the threshold requirement for the creation of connecting edges. The edges are IPF color coded corresponding to the parent austenite orientation of each edge. The lower figure shows the IPF colored prior austenite orientations assigned to each grain after accumulating votes according to Section \ref{sec:compute-parent-or}.

Initially, as shown in Fig.~\ref{fig:variantGraphIterations}(a), the selected grain is only connected to its immediate neighbors by edges formed by pairs of variants. Initially, 22 nodes are connected by a total of 162 edges, implying that multiple variant combinations were found to satisfy the threshold criterion for each pair of nodes. In lath martensite, it is typical that in each grain, the threshold requirement is met by two variants with low misorientation that make up its block structure (V1-V4, V2-V5 and V3-V6 pairing according to Ref. \cite{Morito2003}). In addition, the variant combinations corresponding to an annealing twin for the best fitting candidate orientation typically meet the threshold requirement, as well as its closest variant pairs. This indicates that as many as sixteen edges corresponding to different variant combinations may be potentially found that connect each pair of nodes in the graph.

After 3 iterations, new edges are created to account for second- and third-degree neighbors. The number of edges decreased relative to the number of nodes, as the weaker connections are pruned by the inflation step of the algorithm. Observation of the IPF coloring of the edges indicates that two colors appear to dominate and correspond to twinned prior austenite orientations. After 10 iterations, tenth-order neighbors would nominally be considered for the selected grain. However, this is not possible due to the fact that the edges must be connected by variants that produce a mutually acceptable parent orientation. At this point, it means that all possible edges for the selected grain are found and additional iterations only result in a gradual pruning of weaker edges. Thus, it is clear that the clusters of orientations cannot grow too large even if the inflation parameter is set to 1. The IPF coloring of the edges after 10 iterations indicates that the selected grain would give a good match with either of the twinned prior austenite grains in the immediate vicinity.

As opposed to the grain graph, the variant graph is able to find and preserve the information on every potential prior austenite solution for each original grain in the graph. A simple accumulation of votes as outlined in Section \ref{sec:vote-based-reconstr} is then enough to produce a prior austenite orientation map that is, based on visual observation, morphologically sound when it comes to non-twinned boundaries, as well as being able to reliably satisfy each individual martensite grain with a suitable parent orientation. As shown by the prior austenite orientation map in Fig.~\ref{fig:variantGraphIterations}(c), some ambiguously indexed, small prior austenite grains remain in local regions when twinned prior austenite grains that are in close proximity share martensite variant orientations in the graph.

\begin{figure}
    \centering
    \includegraphics[width=0.5\textwidth]{{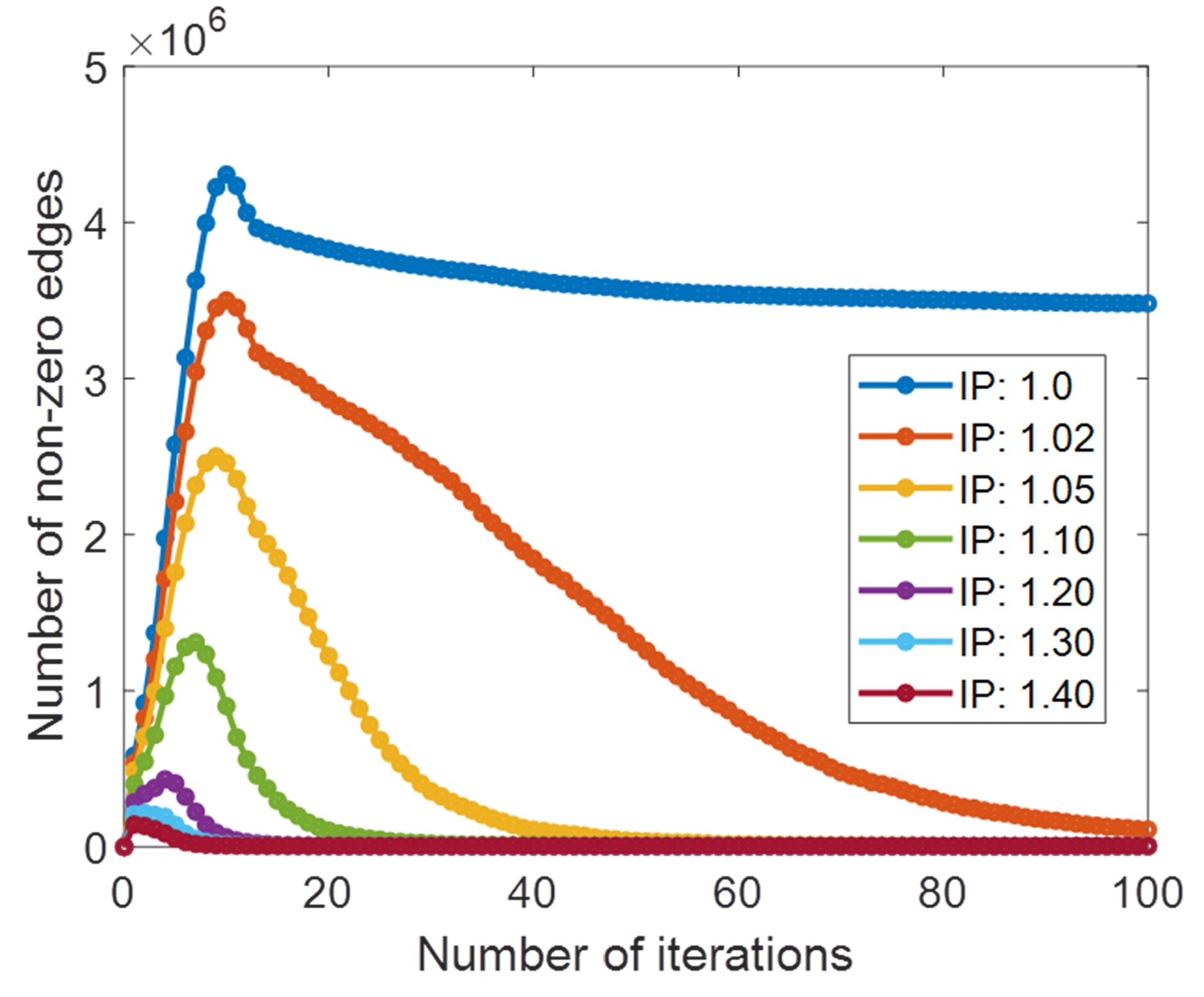}}
    \caption{The convergence behavior of the amount of non-zero edges in the variant graph as a function of inflation parameter.}
    \label{fig:convergence}
\end{figure}

Fig.~\ref{fig:convergence} shows the convergence behavior of the number of non-zero edges at different inflation parameters. As mentioned previously, for an inflation parameter of 1, the convergence of the Markovian graph is not enforced. For this value, the number of edges quickly increases to its maximum up to 11 iterations following which it then slowly decays towards a constant value. This value is much larger than the number of child grains, meaning that the convergence is not towards a single parent variant per child grain but rather, towards a steady state in the computation. 

Slightly increasing the inflation parameter to 1.02 or 1.05 leads to convergence to a low number of non-zero edges. This means the algorithm eventually decides on one possible parent orientation per martensite grain. Increasing the inflation parameter even further to 1.1 and then up to 1.4 leads to extremely rapid convergence for a low maximum value of non-zero edges. Also, the maximum number of non-zero edges is reached after fewer iterations with higher inflation parameter values. This scenario corresponds to a locally restricted search for the best possible parent orientation. 

The range of optimal inflation parameter values, denoting a good trade-off between computational efficiency and an accurate reconstruction that considers many neighboring grains and alternative solutions, is between 1.05 and 1.1. No significant changes in the reconstructed parent microstructure are visible for continued iterations after the maximum number of non-zero edges is reached. Further iterations mostly led to deletion of already unfavorable edges and therefore, did not affect the final result. Stopping the computation when the maximum number of edges are reached is therefore recommended to avoid unnecessary additional iterations and to keep the information on alternative solutions for the best fitting parent orientation.

\section{Discussion}
In this section, a couple of technical details are discussed following which extensions to the variant graph approach are presented. The computational performance of the new, hybrid method is also reported.

\subsection{Pure random walks}
As noted in Section \ref{sec:general-markovian}, the parent orientations are computed even if the Markovian clustering algorithm has not converged. In the specific case that the inflation parameter $\alpha$ is set to $1$, the sum $P_m^i$ of the edge weights $P_{m,n}^{i,j}$ derived in Equ. \eqref{eq:Pmi} are interpreted as a generalization of the voting weights defined in Equ. \eqref{eq:PmiVote}. The major difference is that in conventional local neighbor level voting based parent grain reconstruction, the weight include only first order neighbors while in the variant graph approach, neighbours up to $2^n$ order, where $n$ is the number of iterations, are considered. This explains why the variant graph approach gives better results even for very small $n$. In the corner case of $n=0$, it simply resembles the conventional neighbor level voting based algorithm.

The advantage of running the variant graph algorithm with an inflation parameter $\alpha=1$ and small $n$ is that the resulting voting weight $P_m^i$ is used efficiently to identify child grains when multiple parent orientations are assigned with similar probability. The drawback of setting $\alpha=1$ is that the number of non-zero edges in variant graph increases with the number of iterations. This makes the algorithm unfeasible for large grain maps and many iterations.

\subsection{Diagonal entries} 
The diagonal entries $P_{m,m}^{i,i}$ of the variant graph matrix correspond to self loops of each node of the variant graph. In the grain graph, the weights of these self loops is initialized by $1$, thus making it equally probable to start a random walk at each node. In the case of the variant graph, the weights are initialized to $P_{m,m}^{i,i} = 1/|\mathcal S^c|$, which makes it equally probable to start a random walk with every possible parent orientation variant.

The variant graph also contains so-called pseudo diagonal entries $P_{m,m}^{i,j}$, which correspond to edges connecting different parent variants of the same child grain. Those edges are interpreted as follows: Assuming a certain probability $P_m^i$ for parent orientation variant $\vec g_m^i$ and a second, very similarly oriented parent orientation variant $\vec g_m^j$, then a certain probability for $\vec g_m^j$, namely, $P_m^i \cdot P_{m,m}^{i,j}$ also exists.

Pairs of similarly orientated parent variants appear frequently for experimentally derived and refined irrational orientation relationships that are close to but do not coincide with an ideal rational orientation relationship. 

Although these pseudo diagonal entries possess a physical interpretation, their inclusion in the variant graph is not preferred.

\subsection{Specific morphological conditions}
Up to this point, the initial probability $P_{m,n}^{i,j}$ that two child grains belong to a common parent grain is based solely on the misorientation angle $\omega_p(\vec g_m^i, \vec g_n^j)$ between the potential parent orientations. On the other hand, morphological information may also contribute to this probability. 

As an example, consider the boundary curvature $\kappa$ and denote by $\kappa_{m,n}$ the average over the all boundary segments separating the grains $G_m$ and $G_n$. Following this, the misorientation based weights $P_{m,n}^{i,j}$ are updated according to the curvature by 

\begin{equation}
    \hat P_{m,n}^{i,j} 
    = (1+\beta \kappa_{m,n}) P_{m,n}
\end{equation}

where $\beta$ is a modelling parameter that controls the influence of the curvature to the final weight. The effect of this modification is that only straight boundaries, and not curved boundaries, are more likely to be chosen by the algorithm as parent boundaries. This is especially helpful if the algorithm is presented with an ambiguous situation of finding the correct twin boundary. It is emphasized that any other morphological criterion can also be implemented, as the framework for parent grain reconstruction is a programmatic and fully customizable implementation \cite{Niessen2021b}.

\subsection{Merging variants with small mutual disorientation angles}
\label{sec:mergevariants}
An effective means of reducing memory requirements and computation time is merging closely related variants, i.e., variants that have a small misorientation to each other. For the example data set of austenite to $\alpha'$ martensite transformation, the number of variants are reduced from 24 to 12, which reduces the number of edges by a factor four. This four-fold reduction in number of edges roughly reduces the computation time and memory requirements by a factor of four as well. While it may seem that valuable information about the microstructure is discarded by this step, it is worth noting that child variants with low misorientation to each other are usually paired as neighboring laths and are often not detected during grain reconstruction by the angular threshold criterion in the first place. Inspection of the morphology of the grain map in Fig.~\ref{fig:martensiteOrientations} reveals many irregularly shaped grains, which significantly deviate from a lath-like morphology. It suggests that rather than individual laths, the grains represent blocks comprising multiple variants with a low mutual disorientation angle instead. This observation is made even for a relatively low angular threshold value of 3\textdegree. If it is assumed that variant level precision is lost when the initial grain map is constructed, it follows that the incorporation of full variant level precision into the edges of a variant graph for clustering is simply not necessary. While this assertion definitely bears out in the present and prominently difficult case of $\alpha'$ martensite microstructures in steel, it should be re-examined on other data sets with different parent-child symmetry combinations.

\begin{figure}
    \centering
    \includegraphics[width=1\textwidth]{{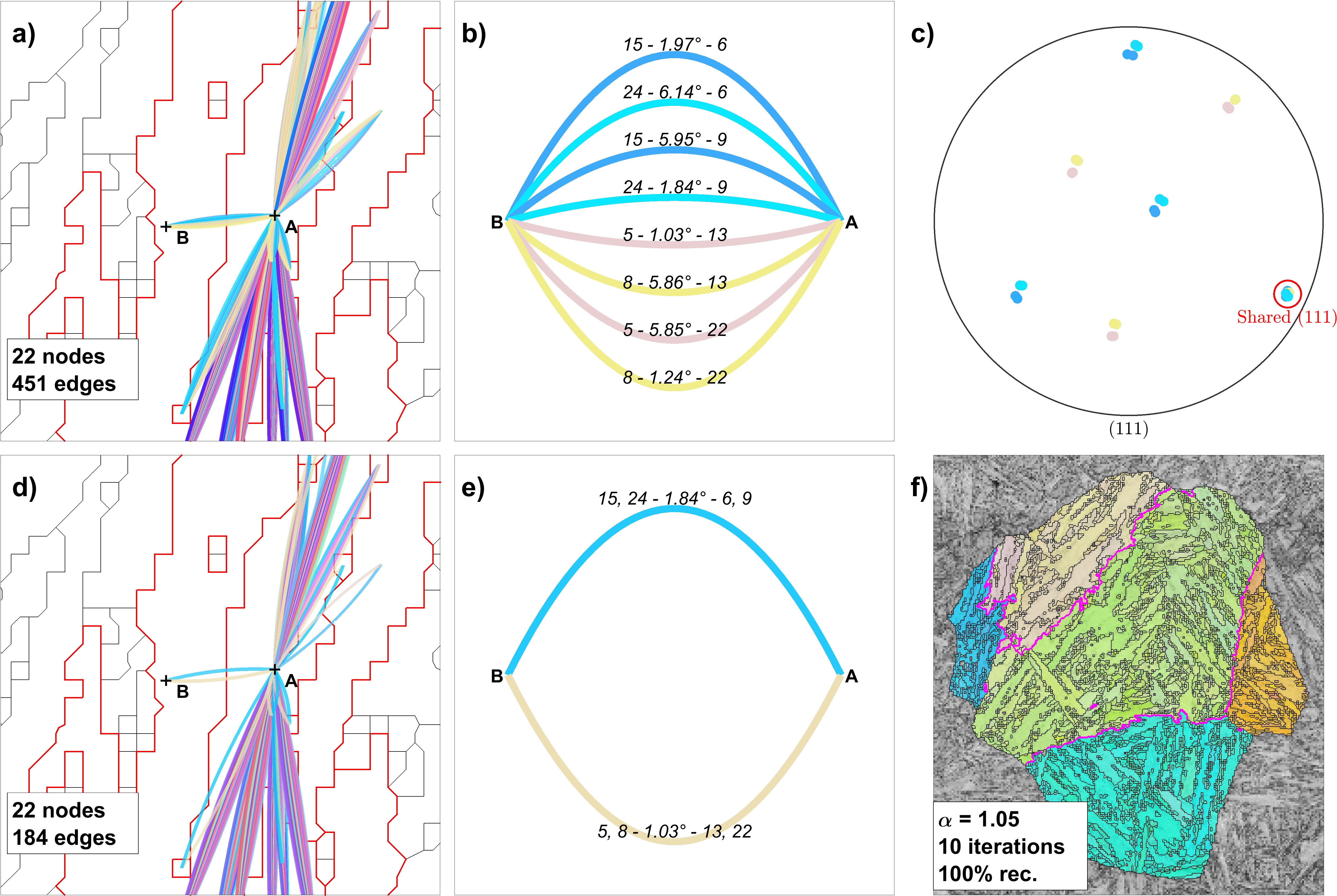}}
    \caption{(a) A small section of the example data set (see Fig. \ref{fig:variantGraphIterations}(a)), showing IPF colored edges according to the mean parent orientation corresponding to the edge. (b) The IPF colored edges between neighboring child grains A and B labelled with the parent variant numbers for the two grains, separated by the angular deviation of the candidate parent variants. (c) (111) pole figure showing the candidate parent variant orientations. The shared (111) plane is highlighted with a red circle. (d) The edges between grain A and its neighboring grain after merging of closely related variants. (e) The number of edges between grains A and B have been reduced from eight to two by the merging process. (f) The final reconstruction result after 10 iterations of clustering following the merging of closely related variants, showing excellent agreement with the parent grain reconstruction from 24 variants in Fig.~\ref{fig:martensiteOrientations}(c).}
    \label{fig:merge_variants}
\end{figure}

Fig. \ref{fig:merge_variants} shows the effect of merging closely related parent variants on the number of edges and the final reconstruction result for the example data set. Using a threshold value $\delta = 8.5^{\circ}$ to determine candidate parent variants, a total of 8 edges are established between grains A and B (Fig. \ref{fig:merge_variants}(a)). Fig. \ref{fig:merge_variants}(b) shows the variant indices and the misorientation angle between the parent variants for each individual edge. Closer inspection reveals that the edges represent closely related parent variants forming two clusters of twin-related parent orientations (see (111) pole figure in Fig. \ref{fig:merge_variants}(c)). In Fig. \ref{fig:merge_variants}(d) and (e), merging closely related parent variants reduces the number of edges between grains A and B from eight to two. In the present case, the lowest misorientation angle between the edges formed by closely related variants represents the strength of the combined edge.

When closely related variants are merged according to this procedure, the resulting reconstruction returns a pair of closely related parent orientations for each child grain. The determination of the final parent orientation therefore requires an additional step to restore variant level precision. The voting algorithm outlined in Section \ref{sec:vote-based-reconstr} is efficient in choosing the locally best fitting parent variant out of the two options determined by the clustering algorithm. Fig. \ref{fig:merge_variants}(f) shows the result of applying this procedure on the variant graph reconstruction result of the example data set. The threshold value for edge detection is $\delta$ = 5\textdegree, and the above merging procedure of closely related parent variants is applied. The inflation parameter is $\alpha$ = 1.05 and clustering is allowed to proceed for 10 iterations. The variant level detail is restored by applying the neighbor level voting algorithm from Section \ref{sec:vote-based-reconstr} to choose the locally best fitting variant from the two closely related variants. Fig. \ref{fig:merge_variants}(f) shows good agreement with the reconstruction using all 24 variants in Fig. \ref{fig:variantGraph}(c).

\subsection{Performance of the variant graph algorithm}
\label{sec:performance}

\begin{figure}
    \centering
    \includegraphics[width=0.75\textwidth]{{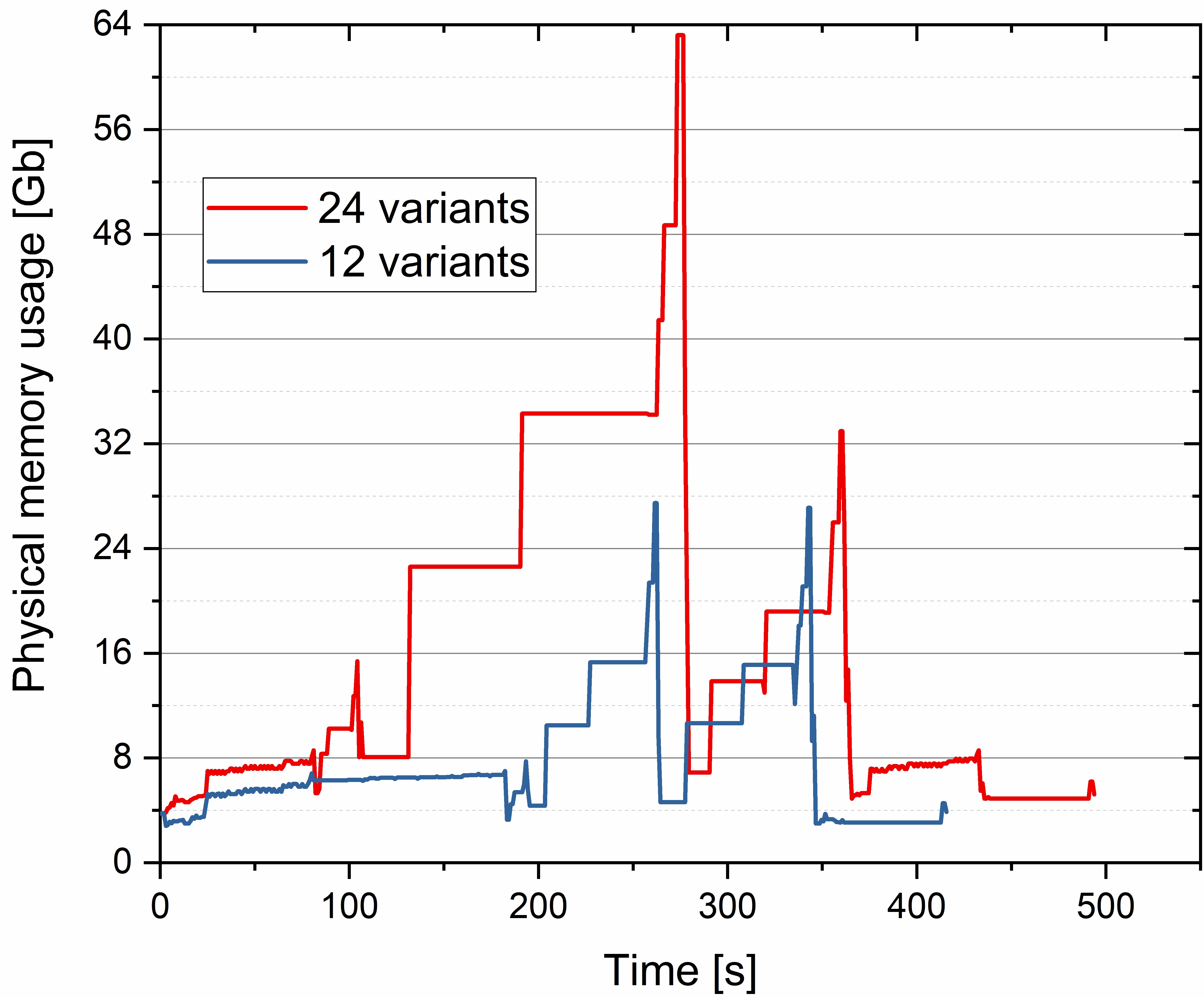}}
    \caption{Comparison of the memory usage vs. time for parent grain reconstruction of the 4 million pixel data set from Fig.~\ref{fig:largeAusteniteMap} using the variant graph algorithm with 24 and 12 variants as discussed in Section \ref{sec:mergevariants}.}
    \label{fig:benchmark}
\end{figure}

\begin{figure}
    \centering
    \includegraphics[width=0.75\textwidth]{{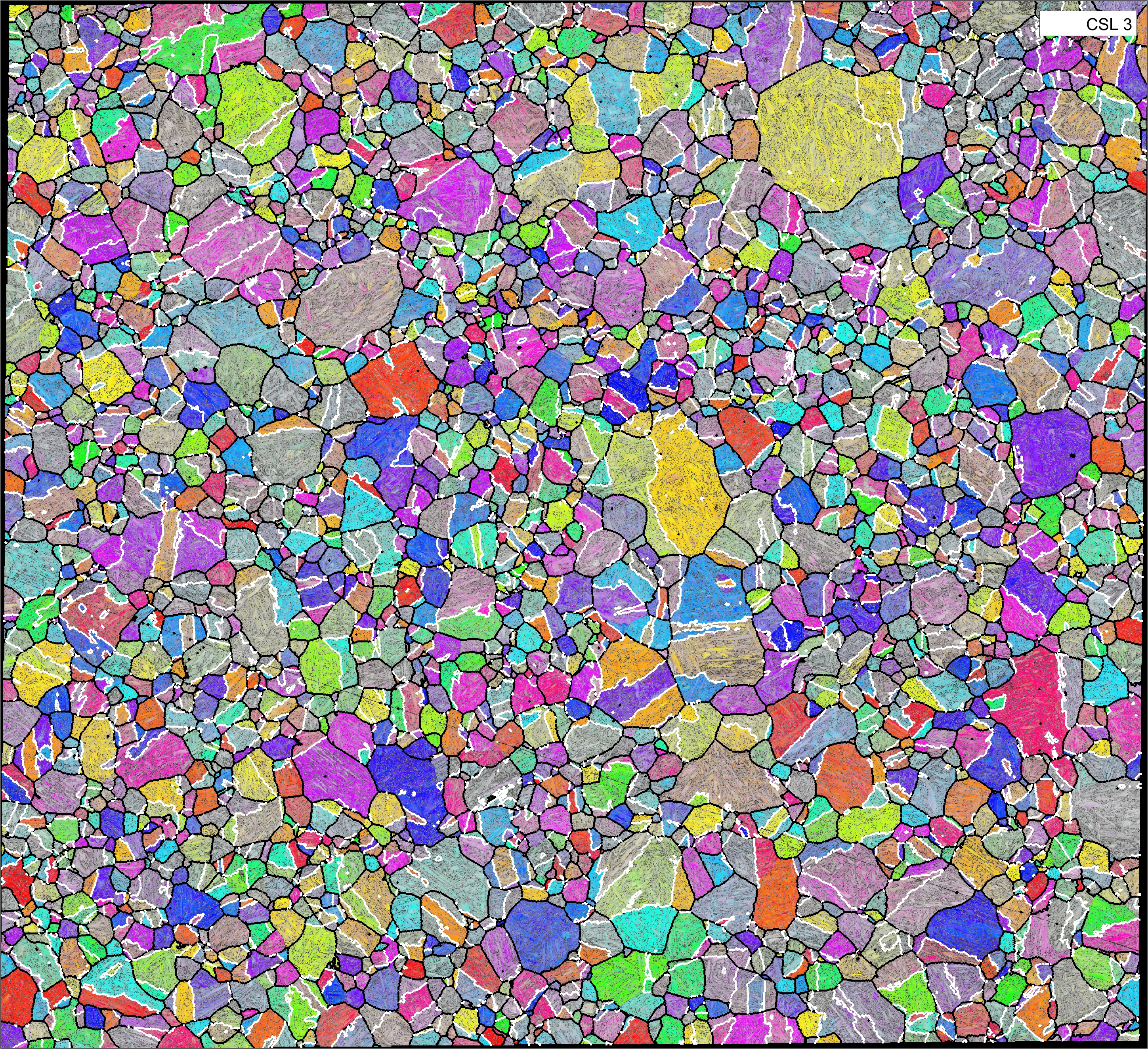}}
    \caption{Reconstructed prior austenite grains with IPF coloring overlaid on the $\alpha'$ martensite band contrast map. $\Sigma$3 boundaries are highlighted in white.}
    \label{fig:largeAusteniteMap}
\end{figure}

Comparing the performance of the grain graph method with the performance of the variant graph method is not straightforward. While the number of edges is significantly smaller in the grain graph, the calculation of parent orientations in the grain graph method requires a separate step and the additional post processing steps are generally non-optional. If the orientation clusters are of only interest, the grain graph algorithm is a quick and robust way of obtaining them. However, as shown in Fig.~\ref{fig:martensiteOrientations}, these clusters rarely correspond to prior austenite grains. With the variant graph approach, the parent orientations are automatically obtained from the graph and only minor post processing is necessary. Running the variant graph algorithm with a large threshold value (as in the current example) yields a good result. However, when working with large data sets, hundreds of millions of edges are created which in turn, translates to high memory usage. In this case, a significant reduction in memory usage is obtained when the extension from Section \ref{sec:mergevariants} is applied which merges variants with mutually close disorientation.

Fig.~\ref{fig:benchmark} shows the memory usage versus time for reconstructing the full example data set, a map of 4 million pixels (see Fig.~\ref{fig:largeAusteniteMap}). The plot compares the performance of the variant graph approach using all 24 variants versus 12 variants as per the adaptation from Section \ref{sec:mergevariants}. As expected, the memory usage is significantly reduced when only 12 variants are used. The 24 variant implementation peaked at 62.2 Gb whereas the 12 variant implementation only required 27.5 Gb at its peak. The 12 variant implementation took 416 s on an Intel(R) Core(TM) i9-10920x processor and was thus 16 \% faster than the 24 variant implementation. 

The amount of memory required is directly proportional to the number of non-zero edges in the graph. The performance metrics presented in Fig.~\ref{fig:benchmark} are just an indicator of the code performance, with all reconstruction parameters kept constant for the different runs. In reality, parameters require tuning for optimal performance with each algorithm. While the code requires a fair amount of memory, its computational performance and efficiency makes it an ideal candidate for virtual memory use when physical memory is insufficient. As a last resort, large data sets could be sectioned and reconstructed in smaller subsets to avoid out-of-memory errors.

Fig.~\ref{fig:largeAusteniteMap} shows the reconstruction of the full example data set, consisting of 4 million data points. The reconstruction was undertaken using the variant graph approach with the extension presented in Section \ref{sec:mergevariants} and an inflation parameter $\alpha$ of 1.05. The reconstructed prior austenite grains are shown in IPF coloring superimposed onto the $\alpha'$ martensite band contrast map with twin boundaries highlighted in white. No post processing was applied.

\section{Conclusions}
\label{sec:conclusions}
This study introduces the new variant graph algorithm for improved parent grain reconstructions from orientation maps of partially or fully phase transformed microstructures. Using the well-known and challenging example of low-carbon lath martensite steel, the algorithm's inherent accuracy in reconstructing parent austenite grains and boundaries is showcased and its computational performance is assessed for a 4 million point data set.  

The variant graph is capable of reconstructing transformation microstructures from any parent-child combination. It is a hybrid algorithm that combines the generalization of the global grain graph with the strengths of the local neighbor level voting based approach. The key advantage of the variant graph is its ability to store all possible parent orientations for each child grain such that following parent grain reconstruction, no additional post processing steps are necessary.

The unique advantages offered by programmatic extensions to the variant graph algorithm include: (i) the ability to account for specific morphological conditions other than the misorientation angle, like boundary curvature, when reconstruction algorithms are faced with ambiguous microstructures, and (ii) the merging of variants with small mutual disorientation angles as an effective means of reducing memory requirements and computation time for parent grain reconstruction.

The variant graph algorithm is implemented as an addition to the phase transformation analysis module in MTEX 5.8 and is freely available for download by the community.

\section{Acknowledgements}
This work is partially financed by the Strategic Innovation Program for Metallic Materials, Vinnova, the Swedish Energy Agency, and Formas as well as the European Union’s Research Fund for Coal and Steel research program under grant agreement number RFCS-2018-847296. Frank Niessen acknowledges the Danish Council for Independent Research (Grants DFF-8027-00009B and DFF-9041-00145B) for financial support.

\bibliography{references}

\end{document}